\documentclass[structabstract]{aa}  
\usepackage{graphicx}
\usepackage{txfonts}
\usepackage{url} 

\usepackage{natbib}
\bibpunct{(}{)}{;}{a}{}{,} % to follow the A&A style
\newcommand{\eg}{{\it e.g.}}
\newcommand{\ie}{{\it i.e.}}

\begin{document}
   \title{Spatially resolved observations of a split-band coronal type-II radio burst}
    \authorrunning{I.~Zimovets et al.}
%   \subtitle{I. Overviewing the $\kappa$-mechanism}

   \author{I.~Zimovets
           \inst{1},
           N.~Vilmer
           \inst{2},
           A.~C.-L.~Chian
           \inst{2,3},
           I.~Sharykin
           \inst{1,4},
           A.~Struminsky
           \inst{1,4}
%          \fnmsep\thanks{Just to show the usage
%          of the elements in the author field}
          }

   \institute{Space Research Institute (IKI) of RAS, Profsoyuznaya Str. 84/32, Moscow 117997, Russia\\
              \email{ivanzim@iki.rssi.ru}
              \and
              LESIA, Observatoire de Paris, CNRS, UPMC, Universit\'{e} Paris-Diderot, 5 place Jules Janssen, 92195 Meudon Cedex,                     France
              \and
              National Institute for Space Research (INPE) and World Institute for Space Environment Research (WISER), P.O. Box 515,                 S\~{a}o Jos\'{e} dos Campos SP 12227-010, Brazil
              \and
              Moscow Institute of Physics and Technology (State University), Institutskii per. 9, Dolgoprudny, Moscow Region, 141700, Russia 
                         }

%  \date{Received September 15, 1996; accepted March 16, 1997}

% \abstract{}{}{}{}{} 
% 5 {} token are mandatory
 
  \abstract
   { The origin of coronal type-II radio bursts and the nature of their band-splitting are still not fully understood, though a number of scenarios were proposed to explain them. This is largely due to the lack of detailed spatially resolved observations of type-II burst sources and of their relations to magnetoplasma structure dynamics in parental active regions. }   
   { To make progress in solving this problem on the basis of one extremely well observed solar eruptive event. }
   { The relative dynamics of multi-thermal eruptive plasmas, observed in detail by the Atmospheric Imaging Assembly onboard the Solar Dynamics Observatory and of the harmonic type-II burst sources, observed by the Nan{\c c}ay Radioheliograph at ten frequencies from 445 to 151~MHz, is studied for the 3 November 2010 event arising from an active region behind the east solar limb. Special attention is given to the band-splitting of the burst. Analysis is supplemented by investigation of coronal hard X-ray (HXR) sources observed by the Reuven Ramaty High-Energy Solar Spectroscopic Imager. }
   { It is found that the flare impulsive phase was accompanied by the formation of a double coronal HXR source, whose upper part coincided with the hot ($T \approx 10$~MK) eruptive plasma blob. The leading edge (LE) of the eruptive plasmas ($T \approx 1-2$~MK) moved upward from the flare region with the speed of $v \approx 900-1400$~km~s$^{-1}$. The type II burst source initially appeared just above the LE apex and moved with the same speed and in the same direction. After $\approx20$~s it started to move about twice faster, but still in the same direction. At any given moment the low frequency component (LFC) source of the splitted type-II burst was situated above the high frequency component (HFC) source, which in turn was situated above the LE. It is also found that at a given frequency the HFC source was located slightly closer to the photosphere than the LFC source.  }
   { Based on the set of established observational facts it is concluded that the shock wave, which could be responsible for the observed type-II radio burst, was initially driven by the multi-temperature eruptive plasmas, but later transformed to a freely propagating blast shock wave. The most preferable interpretation of the type-II burst splitting is that its LFC was emitted from the upstream region of the shock, whereas the HFC -- from the downstream region. The shock wave in this case could be subcritical. }

   \keywords{Sun: corona -- Sun: flares -- Sun: radio radiation -- Sun: X-rays, gamma rays 
               }

   \maketitle
%
%________________________________________________________________

\section{Introduction}
 \label{Introduction}

   It is generally accepted that coronal type-II radio bursts are a signature of MHD shock waves \cite[\eg{},][]{Zheleznyakov70,Wild72,Nelson85,Mann95,Cairns11}. Nevertheless, there is a long-lasting question, whether these shock waves are 1) blast shocks due to explosive flare energy releases or 2) piston shocks driven by eruptive magnetoplasma structures often evolving into coronal mass ejections (CMEs). Numerous observational evidences were reported in favor of both the first and the second scenario \citep{Nelson85,Aurass97,Cliver99,Vrsnak08,Vourlidas09}. To make progress in solving this problem, spatially resolved observations of type-II burst sources at several frequencies in the lower corona ($\lesssim 2R_{\odot}$) are required. Radio observations must be supplemented by high-precision, high-cadence, multi-wavelength observations of magnetoplasma structure dynamics in parental active regions. It is obvious that limb events are best suited for this purpose. 

The partially behind the East limb solar flare of 3 November 2010 (C4.9~class, $\approx$12:10~UT) is a prominent candidate to satisfy these requirements. It was accompanied by a split-band decimetric/metric type-II radio burst, whose sources were well observed by the Nan{\c c}ay Radioheliograph \citep[NRH]{Kerdraon97} in all ten working frequencies from 445 up to $\approx$151 MHz. An important fact is that the radio sources were observed at low altitudes between $\approx$0.25 R$_\odot$ and $\approx$0.65 R$_\odot$. Moreover, the Atmospheric Imaging Assembly \citep[AIA;][]{Lemen11} onboard the Solar Dynamics Observatory observed an erupting multi-thermal magnetoplasma structure in the lower corona $\lesssim$0.4 R$_\odot$ \citep[see][for details]{Reeves11,Foullon11,Cheng11}.

The nature of the band-splitting effect, often observed in coronal and interplanetary type-II bursts, is still an unsolved riddle, although several mechanisms were proposed to explain it \citep[see, \eg{},][as reviews]{Kruger79,Nelson85,Vrsnak01,Cairns11}. The first popular mechanism, initially proposed by \citet{McLean67}, assumes the existence of two (or more) regions with different physical characteristics (say, plasma concentration) along a shock front. The second popular mechanism was proposed by \citet{Smerd74,Smerd75}. It suggests that the two sub-bands of a splitted coronal type-II burst are due to coherent plasma radio emission simultaneously generated in the upstream and downstream regions of a shock wave. The event of 2010 November 3 gives us an opportunity to investigate this interesting and important effect, which in turn can be closely related to the entire problem of the type-II bursts origin. 

It should be noted that combined analysis of this type-II radio burst on the 3rd of November 2010 and associated plasma eruption was reported by \cite{Bain12} very recently. However, \citeauthor{Bain12} mainly concentrated on the dynamics of the most intense type-II burst sources. Our study is more focused on the band-splitting effect. 

The paper is organized as follows. Analysis of the observational data is described in Section~\ref{Observations}. Results of the analysis are summarized in Section~\ref{Results}. These results are discussed and interpreted in Section~\ref{Discussion}. Final remarks are given in Section~\ref{FR}.  

%__________________________________________________________________

\section{Observations}
 \label{Observations}

%%%%%%%%%%%%%%%%%%%%%% FIGURE 1%%%%%%%%%%%%%%%%%% 
   \begin{figure}
   \centering
%   \resizebox{\hsize}{!}{\includegraphics{fig1_colour_cmyk.eps}}
   \resizebox{\hsize}{!}{\includegraphics[bb=67 350 504 787,clip=]{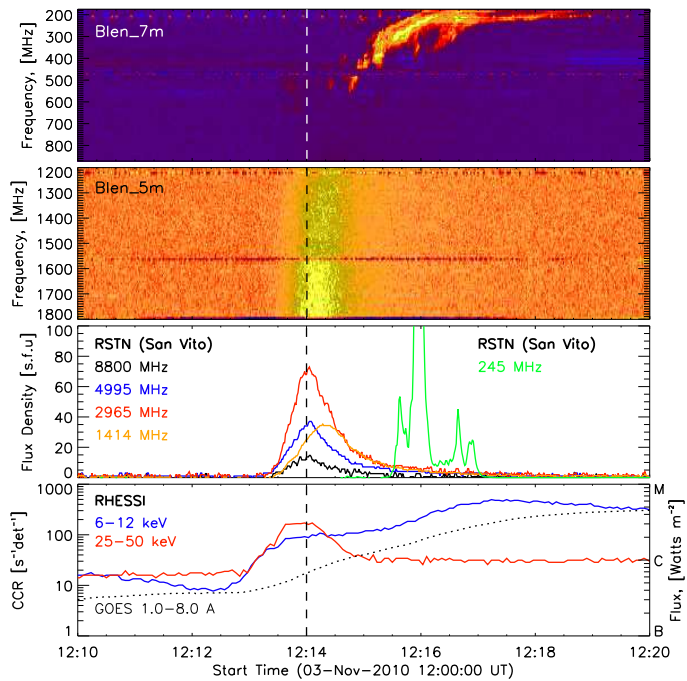}}
%   \includegraphics[width=0.9\textwidth,bb=19 6 564 493,clip=]{fig2.ps}
%   \centerline{\includegraphics[width=0.99\textwidth,bb=19 6 564 493,clip=]{fig1.ps}
%              }
              \caption{ Dynamic radio spectrograms and lightcurves of radio, soft and hard X-ray emissions during the 2010 November 3 eruptive event. Solar radio flux density measured by the telescope in San Vito (RSTN) at 245 MHz is divided by a factor of 10 for clarity. The vertical dashed line indicates the peak time (12:14:00 UT) of the hard X-ray and microwave bursts.                   
                      }
   \label{fig1}
   \end{figure}
%%%%%%%%%%%%%%%%%%%%%%%%%%%%%%%%%%%%%%%%%%%%%%%%%%%%%%%%%%%%%%%%%%%%%%%%%%%%%%%%%%%%%%%%%%%%%%%%%%%%%%%%%%%%%%%%%%%%%%%%%%%%%%%%%%%% 

\subsection{General properties of the event}

According to the Extreme Ultraviolet Imager \citep[EUVI;][]{Wuelser04} onboard the STEREO-B spacecraft, which is located at a point with heliographic coordinates $\approx$S06-E82 during the event, a bright core of the flare appeared at $\approx$S20E97 (\ie{}, $\approx{7^{\circ}}$ behind the east solar limb as seen from the Earth) in the NOAA AR 11121 at $\approx$12:00 UT. This fact will be taken into consideration later, when heights of the investigated objects above the photosphere will be estimated. 

In the frame of the GOES classification the flare was a weak X-ray event of C4.9 class. However, the CCD detectors of EUVI were saturated in the impulsive phase of the flare. Together with the formation of a large eruptive flare loop system this suggests that the flare was actually more powerful than it seemed from the near Earth space, since much of the flare electromagnetic emission was cut off by the limb.

The light curves of the flare radio and X-ray emissions together with radio spectrograms, as observed from the Earth, are shown in Figure~\ref{fig1}. It is seen that the flare impulsive phase started at $\approx$12:13~UT and was accompanied by a burst of microwave and hard X-ray emissions with FWHM$\approx$1~min peaking at $\approx$12:14~UT. The peak of the flare soft X-ray emission observed by GOES-15 was about 6~min later at $\approx$12:20~UT. The flare impulsive phase was also accompanied by groups of decimetric bursts (DCIM; see Figure~\ref{fig4}) as well as a faint type IV burst reported in Solar Geophysical Data between 12:13~UT and 12:17~UT in the 370-864 MHz range. No type-III radio burst was observed during the event. This suggests that accelerated electrons have no access to open field lines. Moreover, no radio emission was observed below $\approx$100~MHz. The pronounced decimetric/metric type-II burst was first observed at 12:14:42~UT, $\approx$45~s after the hard X-ray peak and until $\approx$12:18:00~UT. 

About 20~min after the type-II burst ends, a slow ($v_{linear}\approx241$~km~s$^{-1}$) and rather narrow (angular width $\approx$66$^{\circ}$) CME was first detected by the Large Angle and Spectrometric Coronagraph \citep[LASCO;][]{Brueckner95} at heights $\gtrsim$1R$_{\odot}$ above the photosphere. The CME seemed to be launched from the same NOAA active region 11121 as the investigated flare. The apparent direction of CME motion was similar to those of the flare eruptive plasmas observed by AIA/SDO in the lower corona at heights $\lesssim$0.4R$_\odot$.   

\subsection{Eruptive Plasmas}
 \label{Eruption}

%%%%%%%%%%%%%%%%%%%%%% FIGURE 2 %%%%%%%%%%%%%%%%%% 
   \begin{figure*}
   \centering
%   \resizebox{\hsize}{!}{\includegraphics{fig2_colour_cmyk.eps}}
   \resizebox{\hsize}{!}{\includegraphics[bb=60 383 505 595,clip=]{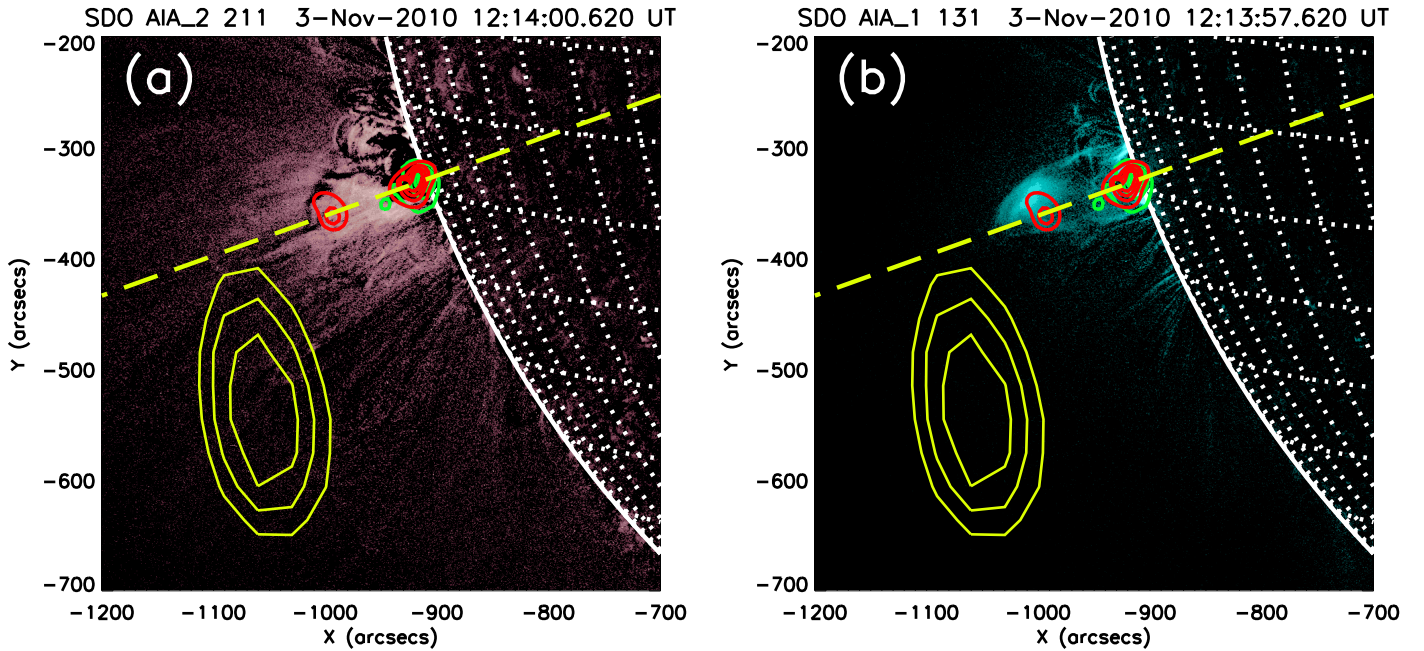}}
%   \includegraphics[width=0.9\textwidth,bb=19 6 564 493,clip=]{fig2.ps}
%   \centerline{\includegraphics[width=0.99\textwidth,bb=19 6 564 493,clip=]{fig1.ps}
%              }
              \caption{ Active area near the eastern limb of the Sun in the impulsive phase of the 2010 November 3 eruptive flare. AIA 211 \r{A} (a) and 131 \r{A} (b) base-difference images are overlaid by the RHESSI 6--12 keV (12:13:54--12:14:14 UT; light green) and 25--50 keV (12:13:54--12:14:14 UT; red) contours (20\%, 40\%, 60\%, 80\% of the peak flux), indicating locations of the flare soft and hard X-ray sources, respectively. AIA 211 \r{A} base image was made at $\approx$12:00:02~UT and 131 \r{A} base image -- at $\approx$12:00:11~UT. Yellow ellipses are the NRH 445 MHz contours (70\%, 80\% and 90\% of the peak flux), which indicate the location of the decimetric radio emission source at the same moment. The thick dashed yellow straight line indicates a projection of the radius-vector passing through the centroid of the flare soft X-ray source onto the image plane.                     
                      }
   \label{fig2}
   \end{figure*}
%%%%%%%%%%%%%%%%%%%%%%%%%%%%%%%%%%%%%%%%%%%%%%%%%%%%%%%%%%%%%%%%%%%%%%%%%%%%%%%%%%%%%%%%%%%%%%%%%%%%%%%%%%%%%%%%%%%%%%%%%%%%%%%%%%%% 

 \subsubsection{General properties}
   
The multi-thermal plasma eruption from the active region was extremely well observed by AIA above the eastern solar limb. Many aspects of these observations were already reported by \citet{Reeves11,Foullon11,Cheng11,Bain12}. Here we just point out a few main observational findings relevant to the problem of the coronal type-II burst's origin, studied in our paper.

First of all, in the impulsive phase of the flare which started at $\approx$12:13~UT, a hot plasma blob (plasmoid-like or flux-rope-like structure) formation and ascent (eruption) was clearly observed in the lower corona ($\lesssim$0.4R$_{\odot}$) in the ``hottest'' AIA channels, centered on the 131\r{A} and 94\r{A} bandpasses \citep[T$\approx$7--11 MK; \eg{},][]{ODwyer10,Lemen11}, (Figure~\ref{fig2}(b)).
 
Secondly, during the hot plasma blob eruption its outer edge seemed to be wraped with an expanding shell (or rim) of relatively cold (henceforth we will call it ``warm'') multithermal plasmas (T$\approx$0.5--2 MK), well observed in 171, 193, and 211\r{A} AIA channels (Figure~\ref{fig3} and Figure~\ref{fig6}). The thickness of this warm shell around the blob, especially above its upper (leading) edge, increased with its rise (this can be clearly seen from Figure~\ref{fig6} and Figure~\ref{fig8}). The process looked like the rising hot plasma blob (most probably, 3D flux rope in reality) pushed up and stretched the overlying magnetic flux tubes, causing plasma to be piled up around the blob. At the same time, some field lines seemed to tear (reconnect) beneath the blob, probably in the quasi-vertical reconnecting current sheet, thus supplying additional heat and magnetic fluxes into the blob.

Another important point is that the apparent direction of the eruptive plasma motion coincided quite well with a projection of the radial direction onto the image plane (Figure~\ref{fig2}, Figure~\ref{fig3}, Figure~\ref{fig6} and Figure~\ref{fig8}). 
  
Finally, morphology of the parental active region, observed by EUVI during the event and by the Helioseismic and Magnetic Imager \citep[HMI;][]{Scherrer11} onboard SDO spacecraft one day later, indicates that the opposite legs of eruptive arcade-like structure were situated at similar helio-latitudes. This means that most probably the eruptive structure was predominantly lying in the plane perpendicular to the image plane if observed from Earth.
  
The observations stated above are informatively brought together in Figure~5 of \citet[]{Cheng11}. Summing up the AIA observations, the entire picture of the event was well consistent with the classical eruptive flare scenario \citep[the so-called CSHKP model;][]{Carmichael64,Sturrock66,Hirayama74,Kopp76}.

\subsubsection{Heights estimation technique}
 \label{Technique}
 
For our further analysis it is of crucial importance to calculate heights of different parts of the eruptive plasmas observed in different moments as accurately as possible. We developed a special technique to calculate the hot plasma blob centroids observed in the 131 \r{A} AIA channel (henceforth ``CE'') and the leading edges (henceforth ``LE'') of the hot and warm plasmas observed in the 131 \r{A}, and 193, 211, and 335 \r{A} AIA channels, respectively.\footnote{ It seems unreasonable to make similar precise data analysis for other AIA channels, because the eruptive plasmas had much more amorphous shapes without clear boundaries in those channels.}

First of all, for the chosen moment and AIA channel a base-difference image is calculated with the pre-flare image at about 12:00:00~UT as a base. Then, the Lee filter is implemented on the calculated base-difference image to remove noise \citep{Lee80}. The Lee filter is an adaptive filter which estimates the local statistics around the chosen pixel. It preserves image sharpness and detail while suppressing noise.

The created base-difference image is divided into a set of columns which are parallel to the choosen direction. Specifically, we choose the radial direction connecting the center of the Sun and the centroid of the soft X-ray source observed by RHESSI in the pre-impulsive phase of the flare (straight dashed line on Figure~\ref{fig2}; see also Subsection~\ref{CorX}). After this, pixel-by-pixel one-dimensional scans are made along each resulting column. The start of each scan is chosen at a point farthest away from the photosphere. Each scan is used to find the farthest point of the eruptive plasma (LE) from the photosphere. Such a point is determined when several consecutive points (\eg{}, N=10, but it is not a critical parameter), if searched from the scan's start, are strictly increasing.

\begin{figure}
   \centering
%   \resizebox{\hsize}{!}{\includegraphics{fig3_colour_cmyk.eps}}
   \resizebox{\hsize}{!}{\includegraphics[bb=75 400 580 785,clip=]{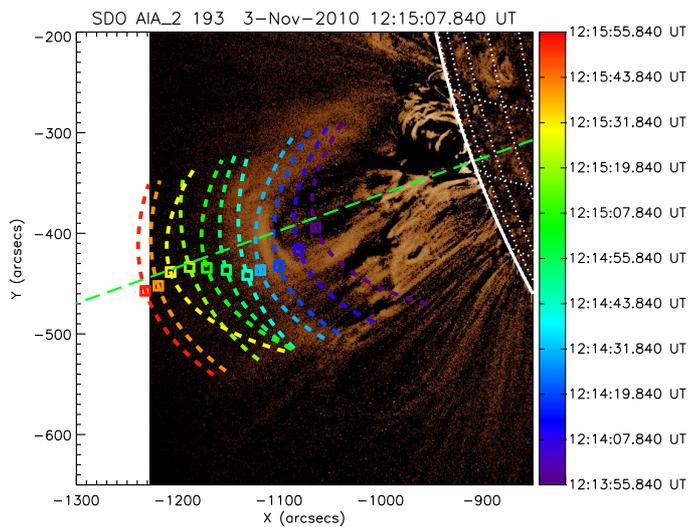}}
%   \includegraphics[width=0.9\textwidth,bb=19 6 564 493,clip=]{fig2.ps}
%   \centerline{\includegraphics[width=0.99\textwidth,bb=19 6 564 493,clip=]{fig2.ps}
%              }
              \caption{ Base-difference image of the active area near the eastern limb of the Sun made by AIA in 193 \r{A} channel at $\approx$12:15:08~UT. The AIA base image was made at $\approx$12:00:08~UT. Dashed parabolic lines of different colours indicate the fitted leading edge of the eruptive plasma at an appropriate moment (see colourbar). The found parabolas' apices are marked by squares. Green dashed straight line indicates a projection of the radius-vector passing through the flare onto the image plane (as in Figure~\ref{fig2}).                     
                      }
   \label{fig3}
   \end{figure}

For the absolute majority of all the analyzed images the points of the LE found in the vicinity of its intersection with the chosen direction are reliably approximated by a parabola (see Figure~\ref{fig3} and Figure~\ref{fig7} as illustrations). Coordinates of the least-square fit parabola's apices are used for the further analysis (particularly they are used in Figure~\ref{fig8}).

Finally, the centroid of the hot plasma blob in each 131 \r{A} image (CE) is determined by averaging over coordinates of the brightness peaks of all one-dimensional scans made in the vicinity of the chosen radial direction.

The technique described above is capable of finding the position of the leading edge of the eruptive plasma (LE) which is likely the leading edge of the magnetoplasma sheath (see Section~\ref{Discussion}). The technique is robust, though it could underestimate slightly ($\sigma\lesssim$15$^{\prime\prime}$) the real position of the LE due to the  restricted AIA sensitivity, the field of view, and the implementation of the noise supressor. The developed technique probably cannot detect the real position of the hypothetical shock wave front. However, if the shock wave front was really formed in the studied event, then, while being in the piston-driven shock wave scenario, it should be situated every time somewhere above the estimated leading edge of eruptive plasmas. The radio observations presented later will show some evidence in favour of this statement.

%%%%%%%%%%%%%%%%%%%%%% FIGURE 3 %%%%%%%%%%%%%%%%%% 
%%%%%%%%%%%%%%%%%%%%%%%%%%%%%%%%%%%%%%%%%%%%%%%%%%%%%%%%%%%%%%%%%%%%%%%%%%%%%%%%%%%%%%%%%%%%%%%%%%%%%%%%%%%%%%%%%%%%%%%%%%%%%%%%%%%% 

\subsection{Coronal X-ray and Decimetric Radio Sources}   
 \label{CorX}           
 
The initial phase of the hot plasma blob ascent (since $\approx$12:13~UT) was accompanied by a single impulsive hard X-ray burst with FWHM $\approx$ 1~min (Figure~\ref{fig1}). This burst was associated with the formation of a double coronal hard X-ray source ($\epsilon_{\gamma} \approx 20-50$~keV) observed by RHESSI (Figure~\ref{fig2}). The lower part of this double source peeped out through the limb and coincided well with the soft X-ray source in the 6-12 keV range, which was situated under the erupting hot plasma blob. At the same time the upper part of the double source seemed to be placed inside the hot erupting plasma blob (Figure~\ref{fig2}). Spectral analysis of the RHESSI data reveals that the hard X-ray emission with $\epsilon_{\gamma} \gtrsim 20$~keV was non-thermally dominated, indicating that it was produced by accelerated electrons. This is also confirmed by spectral analysis of the microwave emission (the detailed analysis will be published elsewhere). After $t \approx$12:15~UT the upper part of the double coronal hard X-ray source disappeared on the RHESSI images, while the lower part and the soft X-ray source retained its position for several minutes. Unfortunately, it is very difficult to study reliably the dynamics of the hard X-ray sources in more detail, because of the low counting rate of the RHESSI detectors at $\epsilon_{\gamma} \gtrsim 20$~keV. Nevertheless, a connection between the coronal hard X-ray sources appearance and plasma eruption is evident. 

The hard X-ray burst was accompanied by an impulsive microwave burst (observed particularly by the San Vito telescope -- a part of the Radio Solar Telescope Network -- RSTN) as well as by groups of decimetric bursts with low and high frequency cut-offs at $f_{l}\approx400$ and $f_{h}\approx700$~MHz, respectively (marked ``DCIM'' in Figure~\ref{fig4}). Source centroids of these decimetric bursts were located $\approx160$~Mm away from the coronal hard X-ray sources and $\approx110$~Mm from a trajectory of erupting plasma (Figure~\ref{fig2}). Note that such a displacement of coronal hard X-ray sources and decimetric bursts has been reported in other events \citep[\eg{},][]{Benz11}. This indicates that non-thermal electrons accelerated in the impulsive phase of the flare, during the eruption of plasma, could be injected into magnetic flux tubes extending to the periphery of the active region. It could be that this phenomenon has similar roots with the type-II precursors reported by \citet{Klassen99,Klassen03}.

\subsection{Type-II burst}
 \label{T2B}

%%%%%%%%%%%%%%%%%%%%%% FIGURE 4 %%%%%%%%%%%%%%%%%% 
   \begin{figure}
   \centering
   \resizebox{\hsize}{!}{\includegraphics{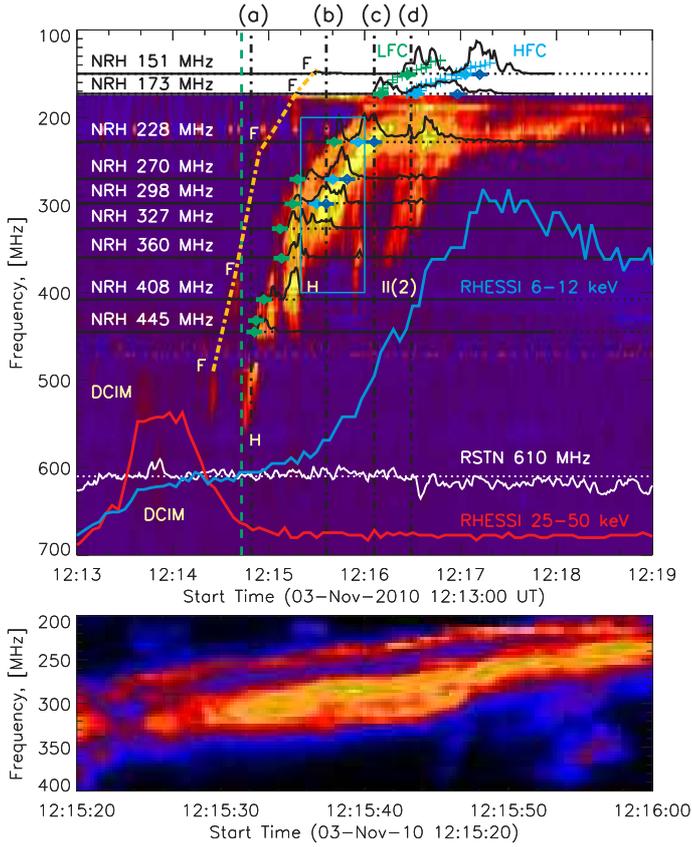}}
%   \resizebox{\hsize}{!}{\includegraphics[bb=10 327 418 830,clip=]{fig4_colour_rgb.ps}}
%    \includegraphics[width=0.7\textwidth,bb=19 6 564 493,clip=]{fig3_aa.ps}
%   \centerline{\includegraphics[width=0.99\textwidth,bb=19 6 564 493,clip=]{fig2.ps}
%              }
              \caption{ Background-subtracted dynamic radio spectrum of the 3 November 2010 solar eruptive event obtained with the Phoenix-4 spectrograph (\textit{upper panel}). Green and turquoise horizontal strokes with diamonds in the middle indicate respectively when the sources of the LFC and HFC of the type-II burst's H-component (H) were observed by the NRH for the first 10~s after its first appearance. Crosses of the same colors represent the LFC's and HFC's first appearence at different frequencies below 180~MHz according to the San Vito spectrograph data. Dark blue horizontal strokes with diamonds indicate the 10 s intervals of the HFC maximal intensity in appropriate frequencies. Time profiles of radio flux density measured by the NRH from the south-eastern sector of the Sun with time cadence of 1~s, as well as 1 second solar radio flux density measured by San Vito telescope (RSTN) at 610~MHz and 4 seconds RHESSI corrected count rates at the 6--12 and 25--50~keV ranges are also depicted. The green vertical dashed line indicates the start time of the type-II burst's H-component. Black vertical dash-dot lines marked by (a), (b), (c), (d) letters on top of the panel indicate four different moments for which four panels of Figure~\ref{fig5} were made. The orange dash-dotted line connects the separated fragments of the type-II burst's F-component (F). The thin turquoise rectangular box indicates a piece of the spectrogram which is represented in the lower panel, using a slightly different color palette. The band-splitting of the type-II burst H-component is clearly seen.                    
                      }
   \label{fig4}
   \end{figure}
%%%%%%%%%%%%%%%%%%%%%%%%%%%%%%%%%%%%%%%%%%%%%%%%%%%%%%%%%%%%%%%%%%%%%%%%%%%%%%%%%%%%%%%%%%%%%%%%%%%%%%%%%%%%%%%%%%%%%%%%%%%%%%%%%%%% 

The background-subtracted dynamic radio spectrum of the Sun, obtained with the Phoenix-4 spectrograph (Bleien, Switzerland; \url{http://soleil.i4ds.ch/solarradio/}) is shown in Figure~\ref{fig4} (upper panel). The bright ($I_{max}\sim10^{3}$~SFU) type-II burst with signatures of herringbone structures started about 45~s after the peak of the hard X-ray and microwave burst. Most probably, we observe mainly a second harmonic emission (denoted by ``H''), whereas the type-II burst at the fundamental frequencies represents only a few minor fragments or wisps (denoted by ``F'' and connected by orange dash-dotted line for clarity). Indeed, often in the events, especially placed far away from the disc center, metric radio emission observed at fundamental frequencies is supressed \citep[\eg{},][]{Zheleznyakov70,Nelson85}. Further we will concentrate mainly on the H-component analysis.

The H-component was first observed at a high frequency of $\approx 561$~MHz at about 12:14:43~UT (marked by green vertical dashed line in Figure~\ref{fig4}). It can be seen that the H-component itself is split into two sub-bands. Henceforth we will call them low and high frequency components (``LFC'' and ``HFC'' in Figure~\ref{fig4}). The less intense frequency band between the LFC and HFC we will call ``band gap''. The band-splitting was more pronounced in the $\approx 200-350$~MHz frequency range (see lower panel of Figure~\ref{fig4}) between $\approx$12:15:20~UT and $\approx$12:16:00~UT. The presence of the LFC and HFC is also confirmed by inspecting the one second time profiles of solar radio flux density measured by the NRH -- they showed two well time-separated flux increases at five NRH frequencies below 298.7~MHz (thick solid black lines on the upper panel of Figure~\ref{fig4}). In general the LFC was less intense (factor of 2) and had more narrower frequency bandwidth (factor of 3--5) than the HFC. In turn, the gap was 2--3 times less intense than the LFC. 

We estimate the mean value of the instantaneous relative bandwidth as $\left\langle \Delta f(t_{i}) / f(t_{i}) \right\rangle = \left\langle [f_{HFC}(t_{i})-f_{LFC}(t_{i})] / f_{LFC}(t_{i}) \right\rangle = 0.16 \pm 0.02$, where $f_{LFC}(t_{i})$ and $f_{HFC}(t_{i})$ is the starting frequency of the LFC and HFC, respectively. The starting frequency is taken each fifth observational moment $t_{i}$ (\ie{}, each 1~s), and the $\left\langle \ldots\right\rangle$-averaging is done over all observational moments $t_{i}$ during the time interval $\approx$12:15:30--12:15:55~UT when the LFC and HFC was best separated on the Phoenix-4 spectrogram. The value of $\left\langle \Delta f / f\right\rangle$ found is consistent with the earlier observations \citep[\eg{},][and references therein]{Nelson85,Mann95}. The LFC and HFC drifting rates ($- df / dt $) ranged from $\approx 1$ to $\approx9$~MHz~s$^{-1}$ with a mean value of $\approx2.2$~MHz~s$^{-1}$. This value is anomalously high in comparison with the previously estimated drifting rates $- df/dt \lesssim 1$~MHz~s$^{-1}$ in the majority of metric type-II bursts \citep[][and references therein]{Nelson85,Mann95}. However, such high values of $\left(- df / dt \right)$ have already been reported a few times for decimetric/metric type-II bursts with high starting frequencies \citep[\eg{},][]{Vrsnak02,Pohjolainen08}.  

It is very fortunate that NRH made observations of the Sun (within its FOV$\approx2^{\circ}\times2^{\circ}$) at all ten operating frequencies -- 445.0, 432.0, 408.0, 360.8, 327.0, 298.7, 270.6, 228.0, 173.2 and 150.9~MHz -- during the November 3, 2010 event. It should also be noted here that the flare time (at around noon in France) is most favorable to make precise observations with NRH during the day, because of minimal zenith distance of the Sun. Implementing the standard technique within the SolarSoftWare to the NRH data obtained with moderate time cadence of 1~s and the half-power beam width (\ie{}, major axis of lobe) of $\approx$45$^{\prime\prime}$, 46$^{\prime\prime}$, 49$^{\prime\prime}$, 56$^{\prime\prime}$, 61$^{\prime\prime}$, 67$^{\prime\prime}$, 74$^{\prime\prime}$, 88$^{\prime\prime}$, 116$^{\prime\prime}$, and 133$^{\prime\prime}$ at frequencies mentioned above, respectively, we generate ten series of 128~pixels $\times$ 128~pixels 2D intensity images of solar radio emission within the entire NRH's FOV$\approx 2^{\circ} \times 2^{\circ} $. The chosen pixel size is $\approx$15$^{\prime\prime}$. In each image the observed type-II burst sources are well fitted by 2D Gaussians that give us estimation of the LFC and HFC source centroid positions. For better statistics at each frequency we find source centroid positions taking an averaging over ten consecutive images at around the moment of interest to us. Specifically, we are most interested in: a) moments of the first appearance of LFC source at each of ten NRH frequencies; b) moments of the HFC source first appearance at five NRH frequencies 298.7, 270.6, 228.0, 173.2 and 150.9~MHz, \ie{}, at those NRH frequencies at which the HFC was clearly separated from the LFC by the band gap according to the Phoenix-4 spectrogram; c) moments of maximum intensity of the HFC at the same five NRH frequencies as in (b).

These three different kinds of moments are marked by green, turquoise and dark blue horizontal strokes with diamonds on the upper panel of Figure~\ref{fig4}, respectively. Lengths of the horizontal strokes indicate ten second time intervals over which averaging of the centroid positions are done at a given frequency.

%%%%%%%%%%%%%%%%%%%%%% FIGURE 5 %%%%%%%%%%%%%%%%%% 
   \begin{figure}
   \centering
%   \resizebox{\hsize}{!}{\includegraphics{fig5_colour_cmyk.eps}}
   \resizebox{\hsize}{!}{\includegraphics[bb=10 5 450 320,clip=]{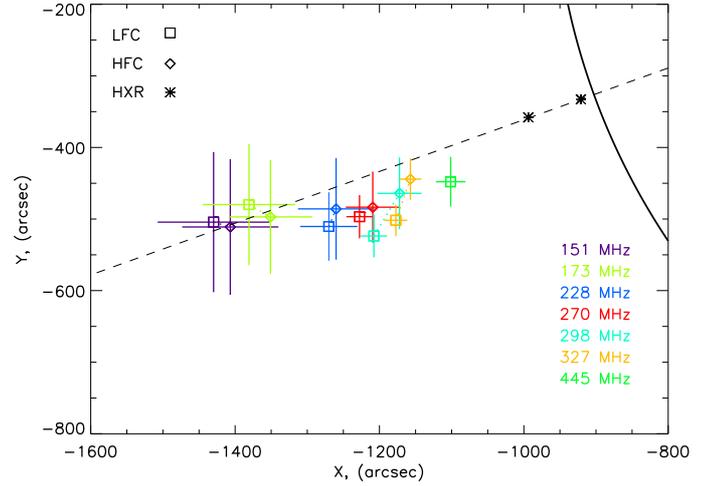}}
%    \includegraphics[width=0.7\textwidth,bb=19 6 564 493,clip=]{fig4_aa.ps}
%   \centerline{\includegraphics[width=0.99\textwidth,bb=19 6 564 493,clip=]{fig4.ps}
%              }
              \caption{ Average positions of the type-II burst's LFC and HFC sources observed by NRH at several frequencies during the 3 November 2010 eruptive event. Error bars of the LFC and HFC sources centroids estimations are shown. Positions of the double coronal hard X-ray source observed by RHESSI in the flare impulsive phase are also plotted by two black asterisks for comparison. The straight black dashed line indicates a projection of the radius-vector passing through the double coronal HXR source onto the image plane. The solar optical limb is represented by black solid arc-like line.                
                      }
   \label{fig5}
   \end{figure}
%%%%%%%%%%%%%%%%%%%%%%%%%%%%%%%%%%%%%%%%%%%%%%%%%%%%%%%%%%%%%%%%%%%%%%%%%%%%%%%%%%%%%%%%%%%%%%%%%%%%%%%%%%%%%%%%%%%%%%%%%%%%%%%%%%%% 

The detailed comparison of the relative positions and dynamics of the LFC and HFC sources and of the eruptive plasmas observed by AIA/SDO will be done in the next subsection. Here we just compare average positions of the LFC and HFC source centroids at six NRH frequencies --- 327.0, 298.7, 270.6, 228.0, 173.2 and 150.9~MHz --- at which the band-splitting of the type-II burst was best seen (see Figure~\ref{fig4}). The averaging is made over the entire duration of the LFC and HFC at a given frequency. Durations of the LFC and HFC are different at different frequencies. Duration of the LFC at 327.0--228.0 MHz is about 20 s, at 173.2 MHz -- 25 s, and at 150.9 MHz -- 40 s. Duration of the HFC at 327.0 MHz is about 16 s, at 298.7 MHz -- 22 s, at 270.6 MHz -- 36 s, at 228.0 MHz -- 31 s, at 173.2 MHz -- 30 s, and at 150.9 MHz -- 40 s. Figure~\ref{fig5} illustrates the comparison. It is seen that at a given frequency the average position of the HFC source centroid was a little bit closer to the photosphere and the flare site than the average position of the LFC source centroid. This subtle effect may be due to large error bars, especially at the lowest NRH frequencies. However as it is systematic at all frequencies, it may be significant and be related to the origin of the band-splitting (see Subsection~\ref{S1} for its discussion). 

It should be briefly noted here how the experimental uncertainties shown in Figure~\ref{fig5} were estimated. The same principle will also be used in the next subsection (\eg{}, vertical bars in Figure~\ref{fig8}). At a given NRH frequency $f_{i}$ the uncertainty is calculated (and most probably it is overestimated) as $\sigma\left(f_{i}\right)=\sigma_{1}\left(f_{i}\right)+\sigma_{2}\left(f_{i}\right)+\sigma_{3}\left(f_{i}\right)$. Here $\sigma_{1}$ is the square root of dispersion of the type-II burst centroid position in the given time interval. $\sigma_{2}$ is half the image pixel diameter, \ie{}, $\sigma_{2}\approx11^{\prime\prime}$ for all $f_{i}$. $\sigma_{3}$ is a measurement of the displacement of the observed radio images  with respect to the true position due to ionospheric waves. These waves with time periods of several tens of minutes are the major contributors to the errors for the absolute pointing of the NRH (A. Kerdraon, private communication). Thus, we estimate $\sigma_{3}$ to be approximately equal to the standard deviation of the centroid positions of a noise storm above the east limb which was observed by the NRH at 173.2 and 150.9~MHz during the 11:00:00--12:13:00~UT time interval just prior to the flare impulsive phase. For the remaining eight frequencies, at which the noise storm was not well observed by the NRH, the errors are estimated using $\sigma_{3}\left(f_{i}\right) \sim f^{-2}_{i}$ rough approximation \citep[][and references therein]{Zheleznyakov70}. The contribution of $\sigma_{3}$ is found to be the largest contributions to the uncertainties, especially at low frequencies.

%%%%%%%%%%%%%%%%%%%%%% FIGURE 6 %%%%%%%%%%%%%%%%%% 
   \begin{figure*}
   \centering
%   \includegraphics[width=0.8\textwidth]{fig6_colour_cmyk.eps}
%   \resizebox{\hsize}{!}{\includegraphics[bb=0 0 455 697,clip=]{fig6_colour_rgb.ps}}
%    \includegraphics[width=0.5\textwidth,bb=0 0 455 697,clip=]{fig6_colour_rgb.ps}
   \centerline{\includegraphics[width=0.8\textwidth,bb=0 0 455 697,clip=]{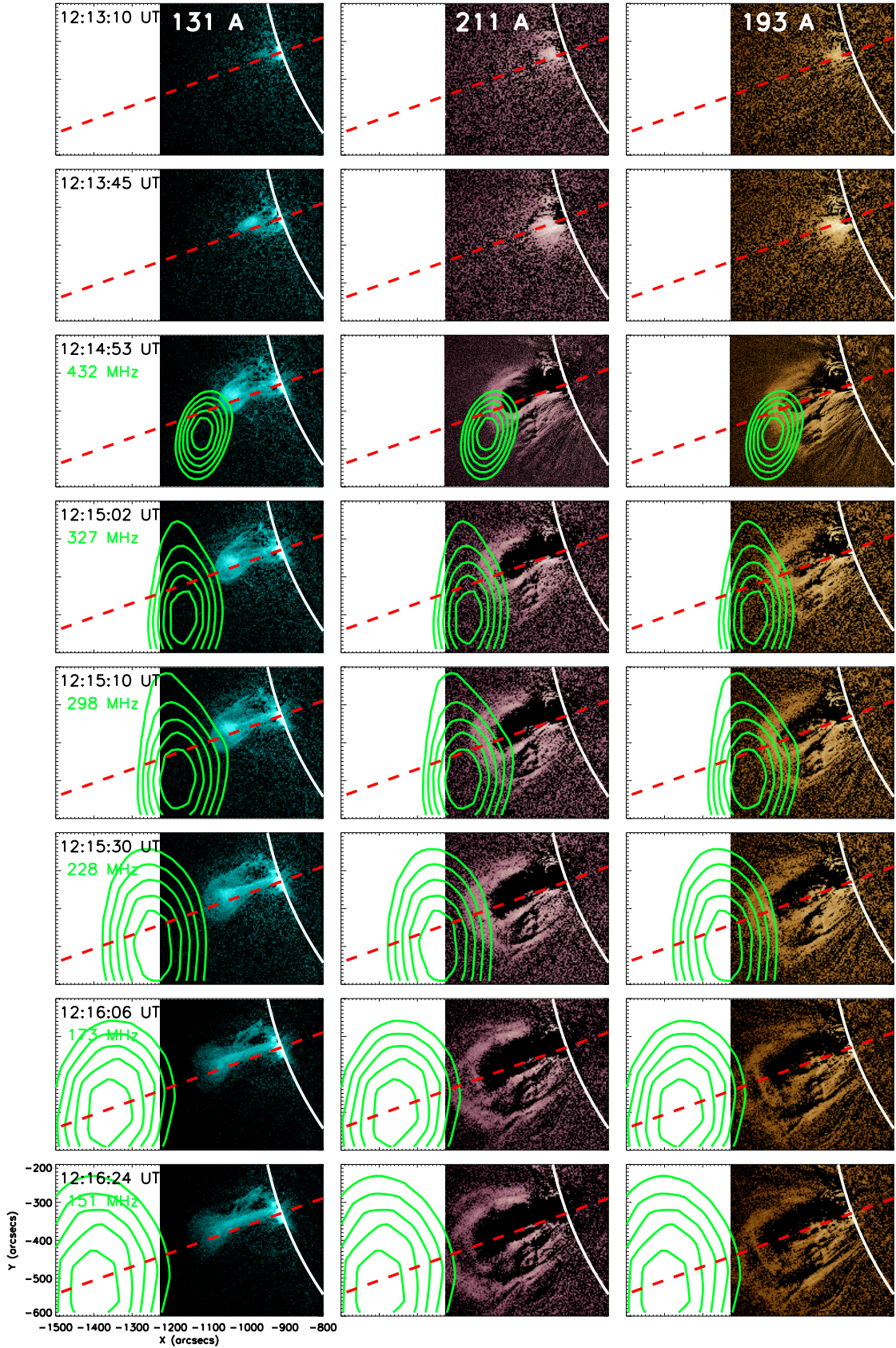}
                 }
              \caption{ Partial time sequence of AIA/SDO 131, 211, and 193~\r{A} base-difference images in between 12:13:10~UT and 12:16:30~UT of 3 November 2010. Overplotted are the iso-intensity contours (50\%, 60\%, 70\%, 80\% and 90\% of the maximum) of the LFC source observed by NRH at different frequencies at a vicinity of times of its first appearance (indicated in the upper left corner of the AIA 131~\r{A} images). One-second integrated NRH data is used. The closest AIA images in time to the NRH ones are shown (the time difference is less than 8 seconds in each case). Solar limb is depicted by the thick white line. The red dashed straight line in all panels indicates a projection of the radius vector passing through the X-ray flare onto the image plane. The AIA's field of view is less than the NRH's one.   
                      }
   \label{fig6}
   \end{figure*}
%%%%%%%%%%%%%%%%%%%%%%%%%%%%%%%%%%%%%%%%%%%%%%%%%%%%%%%%%%%%%%%%%%%%%%%%%%%%%%%%%%%%%%%%%%%%%%%%%%%%%%%%%%%%%%%%%%%%%%%%%%%%%%%%%%%% 

\subsection{Type-II burst sources versus eruptive plasmas}                                     
 \label{Versus}

Figure~\ref{fig6} illustrates the relative dynamics of the multi-temperature eruptive plasmas and of the low frequency component (LFC) source. The images are shown for the time of the first appearance  of the LFC source at a given frequency (these times are marked by green horizontal strokes with diamonds in the upper panel of Figure~\ref{fig4}). Note, that appropriate AIA and NRH images are close to each other within 8~s, \ie{}, they can be considered here as almost simultaneous. It is clearly seen that the LFC source at the highest NRH frequencies initially appeared slightly above the leading edge (LE) of the warm eruptive plasma and that the distance between the LFC source at lower frequencies and the leading edge was increasing with time. This indicates that the agent which excited the LFC source was moving faster than the leading edge of the eruptive plasmas. A similar situation is found for the starting moments of the high frequency component source (HFC) with the only difference being that it was located a little bit closer to the leading edge of the eruptive plasmas.

\begin{figure*}
   \centering
%   \includegraphics[width=0.8\textwidth]{fig7_colour_cmyk.eps}
%   \resizebox{\hsize}{!}{\includegraphics{fig7_colour_cmyk.eps}}
   \includegraphics[width=0.99\textwidth,bb=0 67 454 165,clip=]{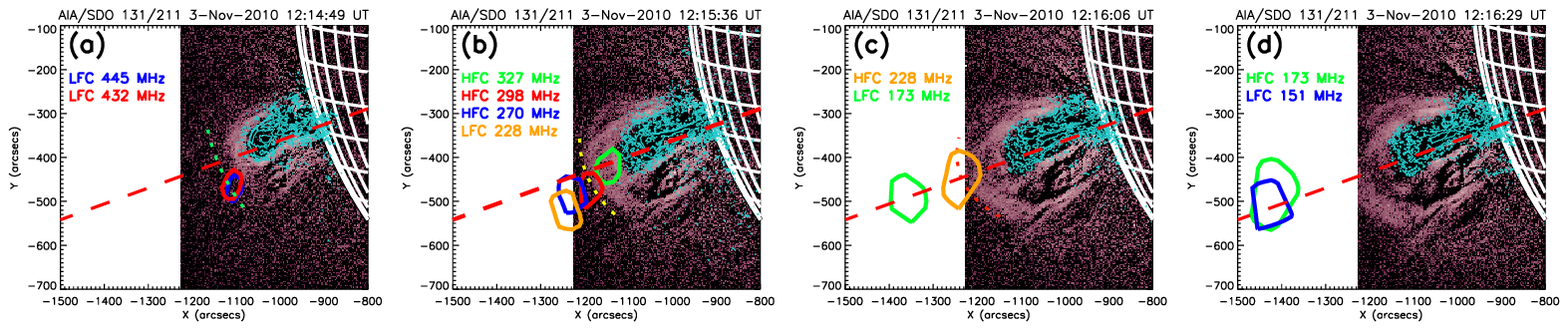}
%   \resizebox{\hsize}{!}{\includegraphics[bb=56 456 507 549,clip=]{fig7_colour_rgb.ps}}
%    \centerline{ \includegraphics[width=0.8\textwidth,bb=6 6 488 396,clip=]{fig4_aa(rgb).ps}
%   \centerline{\includegraphics[width=0.99\textwidth,bb=19 6 564 493,clip=]{fig4.ps}
%              }
              \caption{ Composite base-difference images of the active area near the eastern limb of the Sun made by AIA in 131~\r{A} (turquoise) and 211~\r{A} (purple) passbands at four different times of the 3 November 2010 event. These four times are marked by dash-dotted vertical lines in Figure~\ref{fig4} and Figure~\ref{fig8}. Green, yellow, and red dashed parabolas on panels (a), (b), and (c) respectively indicate the approximated leading edge of eruptive plasma observed by AIA in 211~\r{A} passband. The parabolas' colors are consistent with the colorbar in Figure~\ref{fig3}. Solid lines of different colors are the NRH contours (95\% of the peak flux), which indicate centroids locations of the type-II burst sources at different frequencies (indicated within each panel) at appropriate moments. All AIA and NRH images are matched within 5~s. Red dashed line in all panels indicates a projection of the radius-vector passing through the X-ray flare onto the image plane.                   
                      }
   \label{fig7}
   \end{figure*}

\begin{figure}
   \centering
%   \resizebox{\hsize}{!}{\includegraphics{fig8_colour_cmyk.eps}}
   \resizebox{\hsize}{!}{\includegraphics[bb=84 368 532 818,clip=]{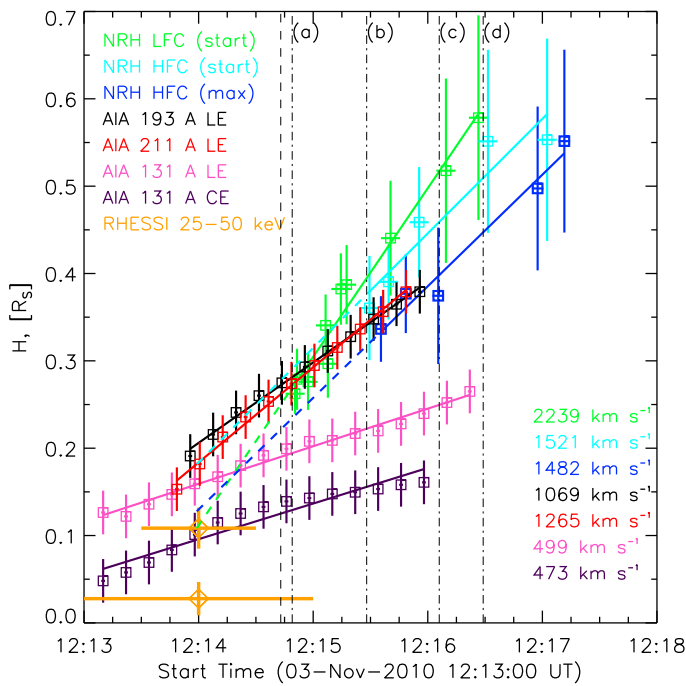}}
%    \includegraphics[width=0.7\textwidth,bb=19 6 564 493,clip=]{fig4_aa.ps}
%   \centerline{\includegraphics[width=0.99\textwidth,bb=19 6 564 493,clip=]{fig4.ps}
%              }
              \caption{ Height-time plot of the type-II burst sources observed by NRH versus different parts of the multi-temperature eruptive plasmas observed with AIA/SDO. The double coronal hard X-ray source heights observed by RHESSI in the flare impulsive phase are also plotted by orange crosses with diamonds. Black dashed and dash-dotted vertical lines indicate the beginning of the type-II burst and four time moments for which panels (a), (b), (c) and (d) of Figure~\ref{fig7} was created, respectively. Error bars of all objects estimations are shown with vertical and horizontal strokes of appropriate colors. Heights of the type-II burst's LFC and HFC sources are given for those time intervals, which are shown by the same colors (green, turquoise and dark blue) in Figure~\ref{fig4}. The least square fittings of the observational data points with the linear functions are plotted by the straight lines of appropriate colors.                  
                      }
   \label{fig8}
   \end{figure}

Figure~\ref{fig7} shows the relative positions of the LFC and HFC sources and of the eruptive plasmas at four different times indicated by vertical dash-dotted black lines and marked by (a), (b), (c) and (d), respectively, in Figure~\ref{fig4} and Figure~\ref{fig8}. It is seen that at any given time both the LFC and HFC sources are located above the leading edge of the erupting plasma. Only the highest-frequency part of the HFC source at 327.0~MHz seems to be located inside the warm eruptive plasma on the panel (b). Further, in Section~\ref{Discussion}, we will discuss that this could be the result of a projection effect. Figure~\ref{fig7} also shows that at a given time the LFC sources are located above the HFC ones. This indicates the presence of a natural plasma density stratification above the erupting plasmas. Both the apex of the leading edge of the erupting plasma and the LFC and HFC source centroids are situated close to the projection of the radius-vector passing through the X-ray flare onto the image plane.

%%%%%%%%%%%%%%%%%%%%%% FIGURE 7 %%%%%%%%%%%%%%%%%% 
%%%%%%%%%%%%%%%%%%%%%%%%%%%%%%%%%%%%%%%%%%%%%%%%%%%%%%%%%%%%%%%%%%%%%%%%%%%%%%%%%%%%%%%%%%%%%%%%%%%%%%%%%%%%%%%%%%%%%%%%%%%%%%%%%%%% 

%%%--- Table 1 ---%%%%%%%%%%%%%%%%%%%%%%%%%%%%%%%%%%%%%%%%%%%%%%%%%%%%%%%%%%%%%%%%%%%%%%%%%%%%%%%%%%%%%%%%%%%%%%%%%%%%%%%%%%%%%%%%%%%%%
\begin{table*}
\caption{ Velocity estimations of different moving objects observed in course of the 3 November 2010 eruptive event.    
}
\label{Table1}
\begin{tabular}{c c c c c c c c}     % define the column alignment
                           % l: left, c: center, r: right
  \hline                   % horizontal line
Object & 131 \r{A} CE & 131 \r{A} LE & 193 \r{A} LE & 211 \r{A} LE & LFC (S) & HFC (S) & HFC (M) \\
  \hline
Velocity, (km s$^{-1}$) & 473$\pm$87 & 499$\pm$72 & 1069$\pm$138 & 1265$\pm$138 & 2239$\pm$150 & 1521$\pm$293 & 1482$\pm$154 \\

  \hline
\end{tabular}
\tablefoot{
CE -- centroid, LE -- leading edge, LFC and HFC -- low and high frequency components of the type-II burst, (S) -- start, \ie{}, the first appearance, (M) -- maximum intensity. 
}
\end{table*}
%%%%%%%%%%%%%%%%%%%%%%%%%%%%%%%%%%%%%%%%%%%%%%%%%%%%%%%%%%%%%%%%%%%%%%%%%%%%%%%%%%%%%%%%%%%%%%%%%%%%%%%%%%%%%%%%%%%%%%%%%%%%%%%%%%%%%%%

To investigate the relative dynamics of the multi-thermal erupting plasmas and of the LFC and HFC sources in more details, we built a height-time plot (Figure~\ref{fig8}), where we present as a function of time the heights above the photosphere of: 
\begin{enumerate}
 \item{ apices of the leading edges (LE) of the warm and hot eruptive plasmas estimated using the technique discussed in Subsection~\ref{Technique}; }
 \item{ centroids of the hot erupting plasma blob (CE); }
 \item{ centroids of the LFC and HFC sources at the moments of their first appearance at different frequencies; }
 \item{ centroids of the HFC sources at the moments of the maximum brightness of the HFC at different frequencies; } 
 \item{ centroids of the double coronal hard X-ray sources. }
\end{enumerate}

Calculated heights of all these objects were corrected for the location of the parental flare region behind the limb. Error bars are indicated in the plot. We also indicate here results of the least square fitting of the observational data-points with a linear function. The linear approximation seems quite reasonable in this particular case. The estimated velocities of the investigated objects are summarized in the lower right corner of Figure~\ref{fig8} and in Table~\ref{Table1}. It shows that the LFC source was moving approximately two times faster than the leading edge of the warm eruptive plasma, which in turn was moving approximately two times faster than the hot plasma blob \citep[in good consistency with the result of][]{Cheng11,Bain12}. The last fact implies that the width of the apparent warm envelope around the hot plasma blob was increasing in time. The HFC source, which has a broader bandwidth than the LFC  at any given time, seems to fill almost all the space between the LFC source and the leading edge (LE) of the warm erupting plasma. It is also worth mentioning here that the approximating curves of the LFC (start) and HFC (maximum) data points (green and dark blue straight lines in Figure~\ref{fig8}, respectively) intersect with the upper part of the double coronal hard X-ray source position in the height-time plot. However, it is difficult to say whether this was purely accidental or not.

All the velocities estimated in Table~\ref{Table1} are probably super-magnetosonic (see also Subsection~\ref{S1} on this issue) at the coronal levels where the type-II burst is observed, \ie{}, at heights $H \approx \left(0.2-0.6\right) \times R_{\odot}$ above the photosphere (this inference will be used further in Section~\ref{Discussion}). Indeed, the sound velocity is estimated as $v_{s} \approx 150-220$~km~s$^{-1}$ if the background temperature is suggested to be of reasonable coronal values $T\approx1-2$~MK. To estimate the Alfv\'{e}n speed we need first to estimate the magnetic field strength $B$. For this purpose we use the rough relation $B=0.5 \times H^{-1.5}$~G, obtained by \citet{Dulk78} from radio observations. It gives $B \approx 1-6$~G. Secondly, we need to estimate the electron plasma density $n_{e}$. This can be done using the entire frequency band of the type-II burst $\approx 560-130$~MHz. Taking into account that the second harmonic emission was observed we infer $n_{e} \approx 5\times10^{7} - 1\times10^{9}$~cm$^{-3}$. Thus, the rough Alfv\'{e}n and fast magnetosonic speed estimation is $v_{A} \approx 310 - 420$~km~s$^{-1}$ and $v_{fms} \approx \sqrt{v^{2}_{s}+v^{2}_{A}} \approx 340 - 470$~km~s$^{-1}$, respectively.

%%%%%%%%%%%%%%%%%%%%%% FIGURE 8 %%%%%%%%%%%%%%%%%% 
%%%%%%%%%%%%%%%%%%%%%%%%%%%%%%%%%%%%%%%%%%%%%%%%%%%%%%%%%%%%%%%%%%%%%%%%%%%%%%%%%%%%%%%%%%%%%%%%%%%%%%%%%%%%%%%%%%%%%%%%%%%%%%%%%%%% 

%%%%%%%%%%%%%%%%%%%%%%% FIGURE 9 %%%%%%%%%%%%%%%%%% 
%   \begin{figure}
%   \centering
%   \resizebox{\hsize}{!}{\includegraphics[bb=0 0 455 697,clip=]{fig9_aa(rgb).ps}}
%%    \includegraphics[width=0.7\textwidth,bb=19 6 564 493,clip=]{fig4_aa.ps}
%%   \centerline{\includegraphics[width=0.99\textwidth,bb=19 6 564 493,clip=]{fig4.ps}
%%              }
%              \caption{ Partial time sequence of AIA/SDO 131, 211, and 193 \r{A} base-difference images in between 12:13:10 UT and 12:17:05 UT of 2010 November 3 overlaid by the iso-intensity contours (90\% and 95\% of the maximum) of the HFC source observed by NRH at different frequencies at a vicinity of times of its first appearance (indicated in the upper left corner of the AIA 131 \r{A} images). 1-second integrated NRH data are used. The closest in time AIA images to the NRH ones are shown (the time difference is less than 8 seconds in each case). Solar limb is depicted by the thick white line. (Note that AIA's field of view is less than that of NRH's one. Red dashed line in all panels indicates a projection of the radius-vector passing through the X-ray flare onto the image plane.              
%                      }
%   \label{fig9}
%   \end{figure}
%%%%%%%%

\section{Results}
 \label{Results}
 
Summary of our findings is presented below.

\begin{enumerate}      
  
  \item{The hot ($T\sim10$~MK) plasma blob started to erupt in an almost radial direction in the impulsive phase of the 3 November 2010 flare. The characteristic velocity of the hot plasma blob upward motion was $v_{CE}\approx500$~km~s$^{-1}$.}
  \item{The hot plasma blob was surrounded by the warm ($T\sim1-2$~MK) expanding rim. The characteristic velocity of the leading edge of this warm rim was $v_{LE}\approx1100$~km~s$^{-1}$, \ie{}, about twice that of $v_{CE}$. }
  \item{The flare impulsive phase was accompanied by the formation of a double coronal hard X-ray source. The lower part of this source seemed to coincide with the near limb legs of the erupting magnetoplasma structure, whereas the upper part was placed somewhere inside the hot erupting plasma blob.} 
  \item{Half a minute after the peak of the flare impulsive phase the type-II radio burst appeared at decimetric/metric wavelengths. Mainly the second harmonic emission was observed in $\approx$ 560--130~MHz range. Signatures of herringbone structures were found but no type-III radio bursts were observed during the entire event. This suggests that accelerated electrons had no access to open magnetic field lines in the course of the plasma eruption.  }
  \item{The type-II burst was splitted in two sub-bands -- low- and high-frequency components (LFC and HFC). The mean value of the instantaneous relative bandwidth was estimated as $\left\langle \Delta f/f \right\rangle = 0.16 \pm 0.02$. This value is within the statistics reported for the split-band coronal type-II radio bursts. }
  \item{The LFC was about 2 times less intense and had 3--5 times more narrow frequency bandwidth than the HFC.}
  \item{The LFC and HFC sources had similar circular shapes, but at a given frequency the LFC source had a slightly smaller size than the HFC one. }
  \item{Initially, the LFC source was observed by the NRH just near the apex of the warm eruptive plasma rim but the LFC source was moving upward at twice the speed of the rim's apex. The characteristic velocity of the LFC source was estimated as $v_{LFC}\approx2200$~km~s$^{-1}$. }
  \item{The apparent direction of the LFC source motion coincided well with the radial one and that of the erupting plasma. }
  \item{At any time the HFC source seemed to fill almost all the space between the LFC source and the leading edge of the warm plasma rim. }
  \item{Linear back-extrapolation of the observational data points on the height-time plot gave evidence that an exciting agent of both the LFC and HFC sources could be launched from the vicinity of the upper part of the double coronal hard X-ray source in the flare impulsive phase. } 
\end{enumerate}

\section{Discussion}
 \label{Discussion}

First of all, it should be noted that according to the AIA/SDO and RHESSI combined observations, the entire picture of the 3 November 2010 event was consistent with the standard eruptive flare scenario, \ie{}, the CSHKP model \citep[see also][]{Reeves11,Foullon11,Cheng11}. Shock waves of various kinds are expected phenomena of such a model \citep[\eg{},][]{Hirayama74,Priest82}. 

However, we shall also clarify here that there are no direct observational evidences of a shock in this event. No shock wave fronts which would have sharply separated the ``disturbed'' (downstream) and ``undisturbed'' (upstream) regions ahead of the leading edge of the erupting plasmas were found using EUV observations of the AIA/SDO. We emphasize that the measured leading edge of the eruptive plasmas is most probably not a hypothetical shock wave front. If a shock wave was indeed formed, its front should be located at each time somewhere above the measured leading edge. Nevertheless, a compelling indirect evidence of a shock wave formation was found in this event with the NRH and Phoenix observations --- the type-II radio burst sources were observed to propagate with (probably) super-magnetosonic speeds above the leading egde of the eruptive plasmas. We will thus assume below that a shock wave was really formed in this event. 

Our discussion below will be limited to the following two subjects: 1) possible origin of the shock wave, 2) possible origin of the observed band-splitting of the type-II radio burst. 

\subsection{Shock wave driver}
 \label{Driver}

As it has already been mentioned in Section~\ref{Introduction}, type-II radio bursts are believed to be produced by MHD shock waves. However, it is still unclear whether these shock waves are 1) bow or piston shocks driven by eruptive magnetoplasma structures or 2) blast shocks due to an explosive flare energy release localised in space and time \citep[see comprehensive reviews on this topic given by, \eg{},][]{Vrsnak00,Vrsnak08}.

\subsubsection{Piston-driven shock wave}
 \label{PDSW}

We found strong evidence in favour of the first (piston-driven shock wave) scenario: 1) the location of the type-II burst sources above the apex of the eruptive plasma leading edge (see  Figure~\ref{fig6}, Figure~\ref{fig7}, Figure~\ref{fig8}), and 2) the same propagation direction of the type-II burst sources and of the erupting plasma apex (see  Figure~\ref{fig3}, Figure~\ref{fig5}, Figure~\ref{fig6}).

These results are similar to the results obtained by \citet{Bain12} for the same event of 3 November 2010 and by \citet{Dauphin06} for the 3 November 2003 flare, when the type-II burst sources were observed originating above the rising soft X-ray loops. Both \citet{Bain12} and \citet{Dauphin06} interpreted their observations in the frame of the piston-driven shock wave scenario. Spatially resolved observations of the coronal type-II burst sources in a close association with the propagating disturbances observed in the soft X-ray range were also reported earlier by, \eg{}, \citet{Klein99,Khan02}, and it was argued that the shock waves, which could be responsible for the type-II bursts, were most probably driven by these disturbances.

The fact that the estimated speed of the agent which excited the low frequency component source (LFC) of the type-II burst was much larger than the speed of the supposed driver ($v_{LFC} \approx 2v_{LE} \approx 2200$~km~s$^{-1}$), \ie{} the erupting plasma, is not contradictory to the piston-driven scenario. The shock front velocity is indeed expected to be equal to the driver velocity only in the case when the driver has a constant geometrical shape, when it propagates in a homogeneous background medium, and also when the oncoming background plasma can wrap the driver's body. A bow shock wave is formed in this case \citep[\eg{},][]{Vrsnak00}. However, in the present event the possible shock driver is propagating in the stratified solar atmosphere with decreasing magnetic field and plasma density. Its geometrical size increases in time (see Figure~\ref{fig6}), and the upstream background plasma can not easily pass the eruptive warm plasma rim because the later seems to remain rooted to the photosphere by magnetic field lines. Thus, the driver looks like a 3D expanding piston (probably not a spherical one but rather flux-rope shaped) which also has a preferred (upward) direction of motion. A situation when a shock wave propagates through the gravitationally stratified corona including dense loops was numerically simulated, \eg{}, by \citet{Pohjolainen08,Pomoell08,Pomoell09}. It was found that under these conditions the shock wave can propagate faster through the corona than its driver (see Subsection~\ref{S1} for further discussion of this issue). It was also found that the shock front is strongest near the leading edge of the erupting plasma.

\subsubsection{Blast shock wave}
 \label{BSW}

Contrary to the piston-driven shock wave scenario, only one piece of indirect evidence (see Result~11 in Section~\ref{Results}) could support the second (blast shock wave) scenario. It implies that a free-propagating blast shock wave could in principle be generated by the impulsive flare energy release which was evidenced by the formation of the coronal hard X-ray sources in the flare impulsive phase. Note, that \citet{Bain12} did not completely rule out this possibility for the present event of 3 November 2010. \citet{Vrsnak06} also reported evidence in favour of the blast shock wave scenario for the 3 November 2003 similar event. On the other hand, Result~11 may be just a coincidence. It alone seems not enough to interpret the studied type-II burst in frame of the blast shock wave hypothesis. Moreover, it is obvious that the blast shock wave hypothesis can not easily explain the same propagation direction of the type-II burst sources and of the erupting plasma apex. For these reasons we conclude that the type-II burst studied here was most probably produced, at least in the initial stage, by a piston-driven shock wave rather than by a blast shock wave.

Additional comments on the assumption of the piston-driven scenario for this event are found below. The value of $v_{LFC}\approx2200$~km~s$^{-1}$ was obtained using the linear least square approximation of all ten observational data points for the LFC sources (shown by green marks on the height-time plot (Figure~\ref{fig8})). However, the first (in time) half of these data points in Figure~\ref{fig8} seems to behave differently than the second half, thus indicating that a linear approximation is probably not the best one. If the linear approximation is performed only on the first group of data points, the slope of this linear fit line is much closer (although a little bit steeper) to the slope of the linear fit for the leading edge of the eruptive plasma (red and black lines on Figure 8). This means that during the beginning of the type-II burst the characteristic velocity of its sources was similar to the velocity of the eruptive plasma leading edge. About 20~s later it became about twice the magnitude, \ie{} $v_{LFC}\approx2200$~km~s$^{-1}$. 

Recent numerical simulations of \citet[][]{Pomoell08,Pomoell09} have shown that the shock wave velocity can quickly exceed the velocity of the driver (flux rope) and then the shock escapes from the driver. In other words, it was found that the shock wave can be of the piston-driven type in the beginning of eruption and after some time it can transform to the freely propagating blast wave. These findings of \citet[][]{Pomoell08,Pomoell09} are very similar to our observations. 

It should also be mentioned here that the discussed change of slope in the type II data points occurred at the time when the band-splitting was first observed. It is possible that these two facts could be related to each other.

%%%%%%%%%%%%%%%%%%%%%%% FIGURE 9 %%%%%%%%%%%%%%%%%% 
   \begin{figure}
   \centering
%   \resizebox{\hsize}{!}{\includegraphics{fig9_bw_cmyk.eps}}
   \resizebox{\hsize}{!}{\includegraphics{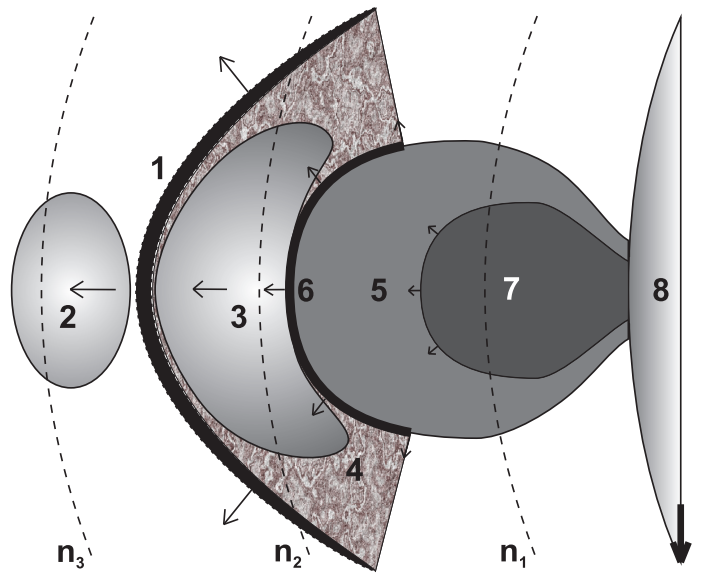}}
%   \resizebox{\hsize}{!}{\includegraphics[bb=0 0 455 697,clip=]{fig9_bw_rgb.ps}}
%    \includegraphics[width=0.7\textwidth,bb=19 6 564 493,clip=]{fig4_aa.ps}
%   \centerline{\includegraphics[width=0.99\textwidth,bb=19 6 564 493,clip=]{fig4.ps}
%              }
              \caption{ Schematic illustration of the 3 November 2010 eruptive event observations combined with their interpretation in the frame of the upstream-downstream scenario (see text). View is from the Heliographic North Pole. Direction to the Earth is marked by a thick black arrow. Notations: (1) hypothetical shock wave, (2) LFC source of the type-II burst, (3) its HFC source, (4) turbulent magnetosheath, (5) warm ($T \approx 1-2$~MK) plasma rim and (6) its leading edge, (7) hot ($T \simeq 10$~MK) erupting flux rope or plasma blob if it is observed from the Earth, (8) photosphere. Thin black arrows show directions of the eruptive plasmas, shock wave, LFC and HFC sources motion. Lengths of the arrows are proportional to the corresponding velocities of motion. Levels of constant undisturbed background electron plasma concentration, assuming the natural gravitational stratification, are marked by black dashed arc-lines, and $n_{1}>n_{2}>n_{3}$.               
                      }
   \label{fig9}
   \end{figure}
%%%%%%%%

\subsection{Split-band effect}
 \label{Splitting}

It is not clear yet, which physical mechanism is responsible for the splitting of type-II bursts' emission. Currently, two main interpretations dominate. This does not mean, of course, that any alternative idea may not be valid \citep[\eg{},][and references therein]{Treumann92,Cairns11}. 

One popular interpretation (henceforth Scenario~1) was proposed by \citet{Smerd74,Smerd75}. It was suggested that the two sub-bands of splitted coronal type-II bursts, the LFC and HFC according to our terminology, could be due to coherent plasma radio emission simultaneously generated ahead of and behind a shock wave front, \ie{}, in the upstream and downstream regions, respectively. 

Another popular interpretation (henceforth Scenario~2), initially proposed by \citet{McLean67}, suggests that different parts of a shock wave front could simultaneously encounter coronal structures of different physical properties, such as electron plasma density or magnetic field. In the case when different parts of a shock wave front propagate through the corona in more than two media with different physical conditions, it is expected to observe a type-II burst with a multiple set of bands. Different situations could occur. For example, those parts of the shock front which are parallel to surfaces of constant electron density would emit more intensively in some narrow frequency ranges than in others. In particular, \citet{McLean67} simulated an idealized situation of a shock front encounter with a streamer and could reproduce split-band of type-II bursts. Similar ideas based on the shock drift acceleration mechanism have been discussed by, \eg{}, \citet{Holman83}. Recently, more sophisticated but ideologically similar numerical experiments of \citet{Knock05} also reproduced splitting of coronal type-II bursts.

\subsubsection{Scenario~1}
 \label{S1}

It seems that our observations geometrically support Scenario~1 more than Scenario~2. The major argument in favour of the upstream-downstream scenario of \citet{Smerd74,Smerd75} is that at every time the low frequency component (LFC) source is  located above the high frequency component (HFC) one, and that the HFC source fills almost all the space between the LFC source and the leading edge of eruptive plasmas (see Figure~\ref{fig7} and Figure~\ref{fig8})\footnote{It should be noted here that the HFC source at 327.0~MHz appears to be located inside the warm eruptive plasma rim in Figure~\ref{fig7}(b). This could be due to projection of the curved HFC source (because of a curved hypothetical shock wave front) into the image plane. See Figure~\ref{fig9} as an illustration.}. This is schematically illustrated in Figure~\ref{fig9}. In this case the LFC source, situated in the upstream region, can be naturally explained in the frame of some standard shock wave theories, \ie{}, the shock drift acceleration mechanism.

\begin{figure}
   \centering
%   \resizebox{\hsize}{!}{\includegraphics{fig10_bw_cmyk.eps}}
   \resizebox{\hsize}{!}{\includegraphics{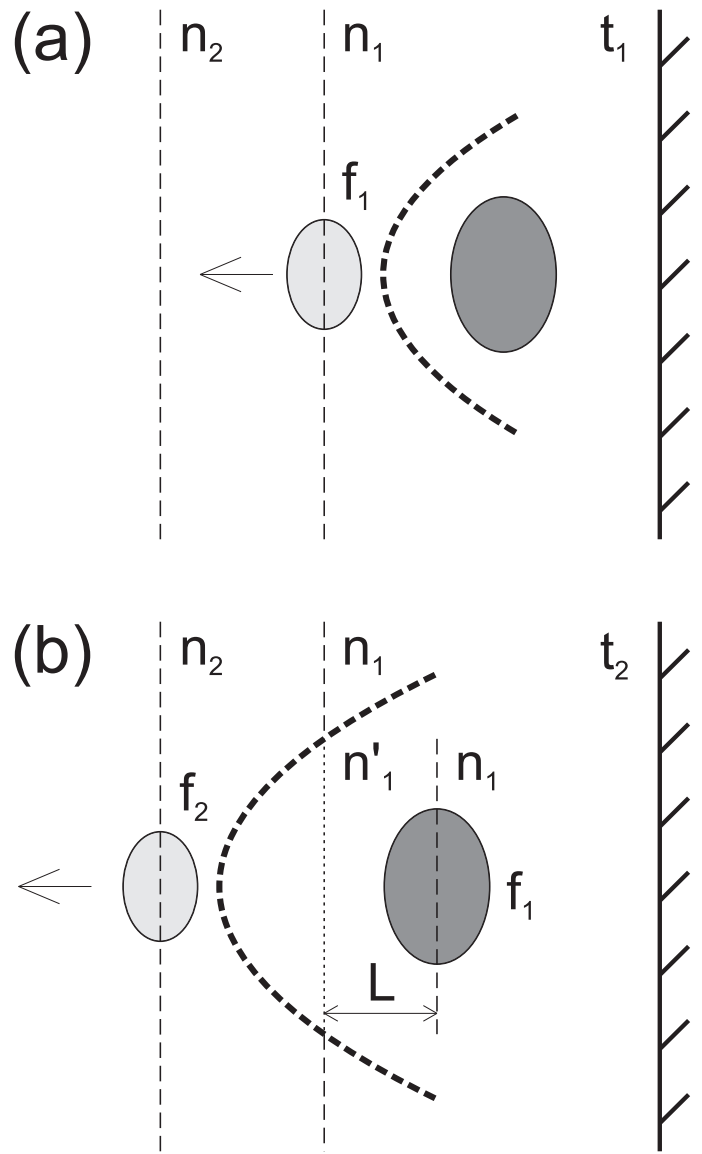}}
%   \resizebox{\hsize}{!}{\includegraphics[bb=0 0 455 697,clip=]{fig9_aa(cmyk).eps}}
%    \includegraphics[width=0.7\textwidth,bb=19 6 564 493,clip=]{fig4_aa.ps}
%   \centerline{\includegraphics[width=0.99\textwidth,bb=19 6 564 493,clip=]{fig4.ps}
%              }
              \caption{ Schematic illustration of the lower location of the HFC sources of the type-II burst with respect to the LFC sources observed at the same frequency. Panel (a) corresponds to an instant $t_{1}$, which is before the instant $t_{2}$ ($t_{2}>t_{1}$) of panel (b). Solar surface is depicted by thick black line with inclined ticks on the right. Thick dashed arc-like line shows the shock wave front. Light and dark grey ellipses represent the LFC and HFC sources, respectively. Horizontal arrow shows the direction of their movement. Levels of constant plasma density are shown by vertical straight dashed and dotted lines. Corresponding plasma densities are marked by $n_{1}$, $n_{1}'$, $n_{2}$, with $n_{1}'>n_{1}$ and $n_{1}>n_{2}$. Corresponding second harmonic of plasma frequencies, at which the LFC and HFC sources are emitted, are marked by $f_{1}$ and $f_{2}$, with $f_{1}>f_{2}$.              
                      }
   \label{fig10}
   \end{figure}

There is another important argument in favour of Scenario~1. It was found (see Subsection~\ref{T2B} and Figure~\ref{fig5}) that at a given frequency the average position of the HFC source centroid was a little bit closer to the photosphere (and the flare site) than the average position of the LFC source at the same frequency. Although this is a subtle effect, it shows that at any time the plasma density in the region, from where the HFC sources are emitted, is enhanced relative to the undisturbed background plasma density. In our opinion, the natural explanation of this effect is that at a given frequency the HFC sources were emitted below the shock wave front --- in the downstream region, \ie{}, in the magnetosheath, whereas the LFC sources were emitted at the same frequency in the upstream region. This idea is schematically illustrated in Figure~\ref{fig10}.     

It can be recalled here that both similar and opposite behaviours of the LFC and HFC sources were reported earlier in the literature \citep[\eg{},][]{Dulk70,Nelson75,Aurass97,Khan02}. In some cases the LFC sources are located farther from the flare site (or from the photosphere) than the HFC sources at the same frequencies but in some other cases they are closer to or almost at the same positions. All these earlier observations (known to us) were made at frequencies below $\approx160$~MHz or for flare regions which were located close to the center of the visible solar disk. This makes it difficult to carry out a direct analogy between these and our observations.

%%%%%%%%%%%%%%%%%%%%%%% FIGURE 10 %%%%%%%%%%%%%%%%%% 
%%%%%%%%

Scenario~1 meets however with a couple of difficulties. The generation of HFC emission requires intense electron plasma waves in the downstream region, whereas in situ measurements near the interplanetary shocks reveal them mainly in the upstream region \citep[\eg{},][]{Bale99,Thejappa00,Hoang07,Pulupa08}. Generation of strong Langmuir turbulence in the downstream region is not easily understood also from the theoretical point of view \citep[\eg{},][]{Treumann92,Cairns11}. Moreover, the type-II radio emission itself is also generally observed from the upstream region of interplanetary shock waves, but not from the downstream region \citep[\eg{},][]{Reiner98,Bale99}. 

There are however observational evidences of radio emission coming from the downstream region of interplanetary shocks \citep[\eg{},][]{Hoang92,Lengyel-Frey92,Moullard01}. In addition: 1) some properties of shock waves in the corona may vary from those in the interplanetary medium, 2) spacecraft measurements in the interplanetary space are still limited both by single point measurements and sensitivity of instrumentation, 3) theory of collisionless shocks is still under development and we still do not fully understand them. Also, some suggestions that a shock front has a wavy (rippled) shape, allow us to explain the downstream populations of energetic electrons even in the frame of the standard shock acceleration theory \citep[\eg{},][]{Vandas00,Lowe00}. Indeed, anisotropic populations of suprathermal electrons were commonly found downstream from those portion of the Earth's bow shock where the shock normal was quasi-perpendicular to the upstream magnetic field, though suprathermal electrons sharply lost their anysotropy and fluxes with increasing penetration into the sheath \citep[\eg{},][]{Gosling89}. This may suggest that populations of nonthermal electrons accelerated at the shock wave front could be also found in the downstream region.

It is not necessary for the non-thermal electron beams responsible for the Langmuir turbulence and the HFC radio emission in the downstream region to be accelerated directly at the shock front. Electrons could also be efficiently accelerated somewhere in the space between the shock front and the leading edge (or on it) of the erupting magnetoplasma structure, \ie{}, in the magnetosheath (Figure~\ref{fig9}). For example, it is known both from in situ measurements \citep[\eg{},][]{Moullard01,Wei03,Gosling07,Wang10,Chian11} and numerical experiments \citep[\eg{},][]{Schmidt03,Wang10} that magnetic reconnection can occur at the interface between the leading edge of interplanetary coronal mass ejections (ICMEs) and background solar wind magnetic field \citep[see also][]{Demoulin08}. Such episodes of magnetic reconnection could supply beams of suprathermal and/or nonthermal energetic electrons to the sheath region \citep{Wang10,Huang12}, thus possibly creating the necessary conditions for generation of the Langmuir turbulence and radio emission there. However, we realize (and emphasize) that this important issue requires further studies.    
       
If the upstream-downstream scenario really takes place in the studied event then it is possible \citep[\eg{},][]{Smerd74,Smerd75,Mann95,Vrsnak02} to estimate the upstream (\ie{}, background) magnetic field ($B_{u}$) using the density jump found at the shock front $X=n_{d}/n_{u} \approx n_{HFC}/n_{LFC}=\left(f_{HFC}/f_{LFC}\right)^2=\left(1+\left\langle \Delta f/f\right\rangle\right)^{2} \approx 1.35$, and compare it with those values which were estimated in Subsection~\ref{Versus} using the formula of \citet{Dulk78}. We will suggest here that the shock wave was oblique rather than purely parallel or perpendicular because a slightly oblique (quasi-perpendicular) shock wave seems to be a more favourable accelerator of electrons in the frame of the shock drift acceleration mechanism \citep[\eg{},][]{Holman83}. For an oblique MHD shock wave \citep[\eg{},][]{Priest82} with an angle $\psi$ between the upstream magnetic field and the shock normal it is possible to derive analytically a quadratic equation $aK^{2}+bK+c=0$ which relates the unknown upstream Alfv\'{e}n-Mach number $M_{A}=\sqrt{K+X}$ with $\psi$, $X$ and the Mach number $M_{S}$. Coefficients of this quadratic equation are defined as $a = 6X/M^{2}_{S}+2\left(X-4\right)\cos^{2}\psi$, $b=X\left(X+5\right)\sin^{2}\psi$, $c=3X^{2}\left(X-1\right)\sin^{2}\psi$. Here it was suggested that the adiabatic index $\gamma=5/3$. 

Firstly, let's estimate the Mach number as $M_{S} \approx v_{LFC} / v_{S}$, where $v_{LFC} \approx 2.2 \times 10^{8}$~cm~s$^{-1}$ is the velocity of the LFC source (see Table~\ref{Table1}) and $v_{S}$ is the upstream sound speed. Within the standard range of coronal temperatures $T \simeq 1-2$~MK we find $M_{S} \approx 10.2-14.5$. Now we can find the physically meaningful solution of the quadratic equation and thus estimate the Alfv\'{e}n-Mach number as $M_{A} \approx 1.06-1.16$ within the entire range of $\psi \in \left(0, \pi/2 \right)$. The upstream magnetic field $B_{u}$ can be estimated as $B_{u} \approx 4.6 \times 10^{-12} v_{LFC} n_{u}^{1/2} / M_{A}$~G, where $n_{u}$ is the upstream electron plasma density in cm$^{-3}$ which was already estimated in Subsection~\ref{Versus} using the observed frequency range of the type-II burst as $n_{u} \approx 5\times 10^{7} - 1\times 10^{9}$~cm$^{-3}$. Thus, we find $B_{u} \approx 6-33$~G and also the plasma parameter $\beta = 2 \left(M_{A}/M_{S} \right)^2 / \gamma \approx 6 \times 10^{-3} - 13 \times 10^{-3}$. These values of $B_{u}$ are about six times larger than those obtained in Subsection~\ref{Versus}. It is not surprising, since the used empirical formula of \citet{Dulk78} is a generalization of observational data of many different active regions.  

A more important inference from the above estimations is the small value of $M_{A} \approx 1.06-1.16$. It indicates that the upstream Alfv\'{e}n speed could be $v_{A} \approx v_{LFC}/M_{A} \approx 1900-2080$~km~s$^{-1}$, that is significantly larger than the observed speeds of the eruptive plasmas (see Table~\ref{Table1}). At first glance this may seemed contradictory to the piston-driven shock wave scenario, which, as it was argued in Subsections~\ref{PDSW} and \ref{BSW}, is preferable in this event. However, numerical simulations of \citet[][]{Pomoell08,Pomoell09} have shown that in the inhomogeneous corona even a sub-Alfv\'{e}nic plasma ejection can launch a shock wave.

It makes sense also to note that the inferred Alfv\'{e}n-Mach number is less than the critical Alfv\'{e}n-Mach number $M_{c}\approx2.76$ for a resistive shock wave \citep[\eg{},][]{Treumann09}. Consequently, the shock wave could be subcritical, regardless of whether it was quasi-parallel or quasi-perpendicular. Electrons were definitely accelerated by the shock wave in the event studied since the type-II burst emission was observed. This gives evidences that subcritical shock waves can accelerate electrons in the corona. This fact could be interesting for theories of charged particle acceleration since it is a supercritical shock wave which is generally believed to accelerate charged particles \citep[\eg{},][]{Mann95,Treumann09}.

\subsubsection{Scenario~2}
 \label{S2}
 
In principle, Scenario~2 could also be implemented in the studied event. One of the possible cases is schematically illustrated in Figure~2(b) of \citet{Holman83}. The efficiency of the type-II radio emission production in the upstream zone by shock drift accelerated electrons depends critically on the angle ($\psi$) between the shock normal at a given point and the upstream magnetic field. Radio emission from the upstream region is then expected only if $\psi$ is restricted to a narrow angular range, within a few degrees of $90^{\circ}$. In the case of a particular mutual arrangement of the shock wave front and of the upstream magnetic field, two separate regions of electron acceleration and thus of enhanced radio emission can be expected. No apparent contradictions between the idea of \citet{Holman83} (especially illustrated in their Figure~2(b)) and our observations are found. If, in the event studied here, the shock wave front was curved rather than plane, the apparent location of the high frequency component sources (HFC) below the low frequency component (LFC) ones could be mainly due to the projection effect. 

Scenario~2 contrarily to Scenario~1 has an important principal drawback --- it can not easily explain correlated intensity and frequency drift variations of the LFC and HFC observed in many type-II bursts \citep[\eg{},][]{Vrsnak01}, as well as in the studied event. It also has difficulties to explain the common range of the relative band-split $\Delta f / f \approx 0.1-0.2$ in many type-II bursts since the solar corona is very inhomogeneous \citep[\eg{},][]{Cairns11}. In our opinion, these facts make Scenario~2 less favourable than Scenario~1.

\subsubsection{An alternative scenario}
 \label{ALI}

\citet[][]{Cairns94} reported observations of fine-structured electromagnetic emissions both from the solar wind and from the Earth's foreshock. It was shown that for fundamental emission, the fine structures above the local plasma frequency $f_{p}$ corresponded to bands separated by near half harmonics of the electron cyclotron frequency $f_{ce}$, \ie, by $n f_{ce} / 2$, where $n$ is a natural number. For harmonic emission the separation was $n f_{ce}$.

In our case the frequency separation for the harmonic emission of the observed type-II burst is $\Delta f \approx 0.16 f_{LFC}$, \ie{}, $\Delta f \approx 30-60$~MHz. The magnetic field estimated in the previous sections is within the range of $B_{1} \approx 6-33$~G (using the upstream-downstream hypothesis) or $B_{2} \approx 1-6$~G \citep[using the formula of][]{Dulk78}. Consequently, the electron cyclotron plasma frequency should be $f_{ce1} \approx 17-92$~MHz in the first case and $f_{ce2} \approx 3-17$~MHz in the second case. The $f_{ce1}$ is in agreement with the observed separation of the LFC and HFC, but $f_{ce2}$ is not. However, the first case corresponds to the situation when the LFC sources were emitted from the upstream region, whereas the HFC sources were emitted from the downstream region. This is in contradiction with \citet[][]{Cairns94} findings, since he reported that the fine-structured emissions were radiated from the upstream region of the Earth's bow shock only.

\section{Final remarks}
 \label{FR}

In this paper the detailed analysis of the partially occulted solar eruptive event of 3 November 2010 was presented. Special attention was given to the search of potential links between the dynamics of eruptive magnetoplasma structure well observed at different temperatures with AIA/SDO, and the sources of the split-band decimetric/metric type-II burst (harmonic emission) observed with the NRH. Simultaneous high precision observations of eruptive structures and spatially resolved observations of decimetric/metric type-II bursts are still rare. This paper deals with the event for which such observations were performed with unprecedented quality. \citet {Bain12} also investigated for this event the origin of the type II burst but we have presented here a more detailed discussion and we also addressed the nature of the type II band-splitting.  

The origin of coronal type-II bursts is still under debate. It is not clear yet whether it is attributed to blast shock waves or to piston shock waves driven by eruptive magnetoplasma structures such as magnetic flux ropes. It is found that the most preferable agent responsible for the coronal type-II burst studied in the paper is the piston shock wave ignited by the eruptive multi-temperature plasmas. The most compelling evidences in favour of this conclusion are 1) location of the type-II burst sources above the apex of the eruptive plasma leading edge, and 2) the same propagation direction of the type-II burst sources and the erupting plasma apex. Since these coupled observational facts can not be easily explained by the blast shock wave, we exclude this possibility. It is also found that at the start of the type-II burst its sources were located just above the apex of the eruptive plasma leading edge and moved with the speed equal to the speed of the eruptive plasma leading edge. But about $20$~s later the speed of the type-II burst sources became twice as large as that of the eruptive plasma leading edge. This indicates that initially the shock wave, which could be responsible for the type-II burst emission, was of a piston type, but later it could transform to a free propagating blast wave. This observation is in close agreement with the results of numerical simulations made recently by \citet[][]{Pomoell08,Pomoell09}.        

Strong observational evidences were found in favour of the hypothesis that the observed type-II burst splitting can be explained by radio sources simultaneously produced upstream and downstream of the shock wave front. The low frequency component (LFC) source was located in the upstream region, whereas the high frequency component (HFC) source was located in the downstream region --- below the shock wave front but above the leading edge of the eruptive plasmas, \ie{}, in the magnetosheath. This can easily explain the slightly lower location of the HFC source relative to the location of the LFC source observed at the same frequency. Based on the band-splitting effect we estimated the Alfv\'{e}n-Mach number as $M_{A} \approx 1.06-1.16$, that is less than the critical Alfv\'{e}n-Mach number. This indicates that in the event studied the shock wave could be subcritical. Nevertheless, due to the fact that the type-II burst emission was observed, we came to the conclusion that even subcritical shock waves could accelerate electrons in the lower corona.

All the conclusions presented above were obtained on the base of only one particular event analysis. To investigate whether they are common or not further investigations of similar well observed events are required. We emphasize here that high-precision, high-cadence, multi-wavelength AIA/SDO observations seem very promising for identification of the direct evidence (\eg{}, jumps of density and/or temperature) of shock waves formation in the lower corona. The first such evidence has already been reported by \citet{Kozarev11} and \citet{Ma11}. Thus, further analysis of joint AIA/SDO and NRH observations of solar events accompanied by decimetric-metric type-II bursts could bring new helpful results on the formation of shock waves in the lower corona and on their physical properties. Such joint observations are most probably rare. In addition, fine-tuned numerical MHD modeling combined with the kinetic simulations of charged particles, are highly required.

\begin{acknowledgements}

The authors are grateful to A.~Klassen, A.~Kerdraon, H.~Bain and S.~Krucker for helpful discussions, to K.~Reeves who initially provided the AIA/SDO data, and to H.~Reid for assistance with the text revision. Courtesy of NASA/SDO and the AIA team. We also thanks to all teams of the instruments (RHESSI, EUVI/SECCHI, GOES, RSTN, Phoenix), whose data are used in the paper. The work was partially supported by the Russian Foundation for Basic Research (project 10-02-01285-a), by the State Contract No.~14.740.11.0086, and by the grant NSch-3200.2010.2. ACLC acknowledges the award of a Marie Curie International Incoming Fellowship and the hospitality of Paris Observatory. NV acknowledges support from the Centre National d'Etudes Spatiales (CNES) and from the French program on Solar Terrestrial Physics (PNST) of CNRS/INSU. The NRH is funded by the French Ministry of Education, the CNES and the R\'{e}gion Centre.

\end{acknowledgements}

\bibliographystyle{aa}

\bibliography{References}

%\begin{thebibliography}{}
%
%  \bibitem[1966]{baker} Baker, N. 1966,
%      in Stellar Evolution,
%      ed.\ R. F. Stein,\& A. G. W. Cameron
%      (Plenum, New York) 333
%
%   \bibitem[1988]{balluch} Balluch, M. 1988,
%      A\&A, 200, 58
%
%   \bibitem[1980]{cox} Cox, J. P. 1980,
%      Theory of Stellar Pulsation
%      (Princeton University Press, Princeton) 165
%
%   \bibitem[1969]{cox69} Cox, A. N.,\& Stewart, J. N. 1969,
%      Academia Nauk, Scientific Information 15, 1
%
%   \bibitem[1980]{mizuno} Mizuno H. 1980,
%      Prog. Theor. Phys., 64, 544
%   
%   \bibitem[1987]{tscharnuter} Tscharnuter W. M. 1987,
%      A\&A, 188, 55
%  
%   \bibitem[1992]{terlevich} Terlevich, R. 1992, in ASP Conf. Ser. 31, 
%      Relationships between Active Galactic Nuclei and Starburst Galaxies, 
%      ed. A. V. Filippenko, 13
%
%   \bibitem[1980a]{yorke80a} Yorke, H. W. 1980a,
%      A\&A, 86, 286
%
%   \bibitem[1997]{zheng} Zheng, W., Davidsen, A. F., Tytler, D. \& Kriss, G. A.
%      1997, preprint
%\end{thebibliography}

\end{document}